\documentclass[%
 reprint,
 amsmath,amssymb,
 aps,
]{revtex4-1}

\usepackage{graphicx}
\usepackage{dcolumn}
\usepackage{bm}

\usepackage{color}
\usepackage{ulem}

\begin{document}

\preprint{APS/123-QED}

\title{Viable dark matter via radiative symmetry breaking in a scalar singlet
   Higgs portal extension of the standard model}
\author{T.G.~Steele}
\author{Zhi-Wei Wang}%
\affiliation{%
Department of Physics and
Engineering Physics, University of Saskatchewan, Saskatoon, SK,
S7N 5E2, Canada
}%
\author{D.~Contreras}
\author{R.B.~Mann}%
\affiliation{%
Department of Physics, University of Waterloo, Waterloo, ON, N2L 3G1, Canada
}%

\begin{abstract}
We consider generation of dark matter mass via radiative electroweak symmetry breaking in an extension of the conformal Standard Model containing a singlet scalar field with a Higgs portal interaction. Generating the mass from  a sequential process of radiative electroweak symmetry breaking followed by a conventional Higgs mechanism can account for less than 35\% of the cosmological dark matter abundance for dark matter mass  $M_s>80\,{\rm GeV}$.  However in a dynamical approach where both Higgs and scalar singlet masses are generated via radiative electroweak symmetry breaking we obtain  much higher levels of dark matter abundance. At one-loop level we find 
abundances of 10\%--100\% with $106\,{\rm GeV}<M_s<120\,{\rm GeV}$. However, when the higher-order effects needed for consistency with a $125\,{\rm GeV}$ Higgs mass are estimated, the abundance becomes 10\%--80\% for $80\,{\rm GeV}<M_s<96\,{\rm GeV}$, representing a significant decrease in the dark matter mass. The dynamical approach also predicts a small scalar-singlet self-coupling, providing a natural explanation for the astrophysical observations that place upper bounds on dark matter self-interaction. The predictions in all three approaches are within the $M_s>80\,{\rm GeV}$ detection region of the next generation XENON experiment.
\end{abstract}
\maketitle


One of the most important outstanding challenges in  physics is to reveal the underlying nature of dark matter. Amongst the numerous proposed dark matter candidates, the singlet scalar extension of the Standard Model is conceptually appealing and has been the subject of much investigation \cite{zee,mcd,cliff,mura,frank,Meissner,Rainer,Cline:ab} (see Ref.~\cite{cliff} for a clear and detailed discussion). This model was first introduced by Silverira and Zee \cite{zee} and then generalized to a complex scalar by McDonald \cite{mcd}. More detailed analyses that included nuclear recoil detection and implications for collider experiments were subsequently studied \cite{cliff}, along with the electroweak phase transition of this singlet extension of the Standard Model \cite{Meissner,Cline:ab}. 
Because it consists of one scalar singlet beyond the Standard Model, it is one of the simplest scenarios for nonbaryonic dark matter. However, it is complicated enough to offer rich properties, such as dark matter stability,  because the Standard Model gauge singlet does not interact with ordinary matter except through the Higgs field (i.e., Higgs portal interactions \cite{Rainer}).
In these models, the stability of dark matter is protected by a scalar singlet $Z_2$ symmetry that prohibits the dark-Higgs-Higgs decay process.

Versions of singlet scalar  models with classical conformal symmetry are particularly interesting as a means for addressing the hierarchy and fine-tuning problems \cite{Susskind,Bardeen} associated with the conventional Higgs mechanism. Classical scale invariance provides a custodial symmetry for Higgs loop corrections \cite{Bardeen,'tHooft:1979bh}, and similar to dimensional transmutation in QCD, leads to 
natural scale hierarchies in a unification context  \cite{Weinberg:1978ym} (see also Ref.~\cite{Hill:2014mqa} for a recent discussion). In these scalar-singlet models, radiative symmetry breaking (i.e., the Coleman-Weinberg mechanism \cite{Coleman:1973jx}) in the hidden (dark) sector gets communicated to the electroweak sector via the Higgs portal interaction \cite{a,b,c,Foot:2010av,Khoze:2013oga,Englert:2013gz,Chang:2007ki,Carone:2013wla,AlexanderNunneley:2010nw}. Typically this requires a vacuum expectation value (VEV) for the scalar singlet field which breaks the $Z_2$ symmetry, and hence additional mechanisms are needed to incorporate dark matter (e.g., mirror dark matter \cite{Foot:2010av}, CP symmetry protected dark matter \cite{Farzinnia:2013pga,Gabrielli:2013hma}, inert doublet dark matter \cite{Hambye:2007vf} and Majorana Dark matter \cite{Heikinheimo:2013cua}). 
 
In this article we take the approach of radiative electroweak symmetry breaking in the Standard Model sector, and explore its implications for the scalar singlet (dark matter) sector. Decays of the dark matter field are protected by  $Z_2$ symmetry without introducing any extra mechanisms. 
In the Coleman-Weinberg mechanism  \cite{Coleman:1973jx}, the VEV for the Higgs is radiatively generated directly in the Higgs sector. The small Higgs coupling Coleman-Weinberg solution \cite{Coleman:1973jx}  is destabilized by top-quark Yukawa contributions, but a large Higgs-self-coupling solution exists \cite{Elias:2003zm,Elias:2003xp}. Recently it has been shown that the 125 GeV Higgs mass observed by LHC \cite{2012gk,2012gu} can be described by radiative electroweak symmetry breaking in the large Higgs self coupling regime \cite{Wang}. The purpose of this article is to show that  radiative symmetry breaking in the large Higgs self-coupling perturbative regime can dynamically generate a dark matter (scalar singlet) mass in a Higgs-portal extension of the Standard Model that  provides a significant proportion of the dark matter abundance. The resulting dark matter mass and the corresponding dark-Higgs coupling are within the parameter space that will be probed by the next generation XENON experiment.

The dark matter scalar singlet extension of the conformal Standard Model has the following scalar sector \cite{zee,mcd,cliff,mura,frank,Meissner,Rainer,Cline:ab}:
\begin{equation}
L=\frac{1}{2}\partial_\mu H\partial^\mu H+\frac{1}{2}\partial_\mu S\partial^\mu S-\frac{k}{2}S^2H^{\dagger}H-\frac{h}{4!}S^4-\lambda(H^{\dagger}H)^2\,,
\label{lagrangian}
\end{equation}
where $H$ is the Higgs field and $S$ is the dark matter (real scalar) singlet field which has no interactions with other Standard Model fields except via the Higgs portal interaction.
As discussed in Ref.~\cite{cliff}, from an effective field theory perspective the absence of higher-dimensional non-renormalizable terms in Eq.~\eqref{lagrangian} assumes other beyond-Standard-Model particles are much heavier than the electroweak scale. Because we are interested in Standard-Model extensions that are conformal at tree level, there are no quadratic terms for the Higgs and dark scalar and there are no $S^3$, $SH^\dagger H$, or similar terms that violate  the $Z_2$ ($S\rightarrow -S$) symmetry. The stability of dark matter is protected by  assuming the $Z_2$ symmetry is unbroken, precluding $SH^{\dagger}H$ terms in \eqref{lagrangian} induced by $\langle S\rangle\ne 0$, and thereby preventing dark matter from decaying through the Higgs portal. With zero VEV, the dark matter field can enter radiative symmetry breaking in two ways: either $S$ is on an equal footing with all other Standard-Model non-Higgs fields, or else both Higgs and scalar singlet masses are radiatively generated.

In the first case, $S$   influences the Higgs effective potential via the Higgs portal interaction. Radiative symmetry breaking first generates the Higgs VEV and then the dark matter gains its mass through the conventional Higgs mechanism via the the dark matter-Higgs coupling $\frac{k}{2}S^2H^{\dagger}H|_{H\rightarrow v}=\frac{k}{2}v^2S^2$. 
Thus in this scenario we consider the effective potential of the Higgs field, which can be rewritten as  $O(4)$ symmetric massless $\lambda\phi^4$ theory because the gauge couplings and top quark Yukawa coupling effects are numerically small in the large Higgs self-coupling regime of interest \cite{Elias:2003xp}.
The effective potential in Coleman-Weinberg (CW) renormalization scheme then has the form \cite{Coleman:1973jx,Jackiw:1974cv}:
\begin{eqnarray}
    V(\lambda,\Phi,\mu) & = & \sum_{n=0}^\infty \sum_{m=0}^n \lambda^{n+1}T_{nm}L^m\Phi^4
\label{effective potential}
\end{eqnarray}
where $L=\log\left(\Phi^2/\mu^2\right)$, $H^\dagger H=\Phi^2=\sum_{i=1}^4\phi_i^2$, and $\mu$ is the renormalization scale which connects the $O(4)$ theory to the Standard Model when $\mu$ equals the electroweak scale $v=246\,{\rm GeV}$. The summation includes leading logarithm $(LL)$, next-to-leading logarithm $(NLL)$, next-to-next-to-leading logarithm $N^2LL$, and in general $N^nLL$ terms.
For the leading logarithm summation we 
obtain
\begin{equation}
V_{LL}=\sum_{m=0}^{\infty}T_{m\, m}\lambda^{m+1}L^m\Phi^4\,.
\end{equation}
Generalizing to the multi-coupling case, assuming the two couplings are $\lambda,k$, the leading logarithm contribution to $V$ can be written as
\begin{equation}
V_{LL}=\sum_{m=0}^{\infty}\sum_{r=0}^{m}T_{m-r+1,\, r}\lambda^{m-r+1}k^{r}L^m\Phi^4\,.
\end{equation}
The form of the multi-coupling case can be further extended to additional couplings for the dark singlet extension model. 
Because of  the Coleman-Weinberg renormalization condition  \cite{Coleman:1973jx,Jackiw:1974cv} 
\begin{equation}
\frac{d^4V}{d\Phi^4}\bigg|_{\Phi=\mu}=24\pi^2 y\,,~y=\lambda/\pi^2\,,
\label{cw_condition}
\end{equation}
it is only necessary to consider terms up to order $L^4$ in the effective potential to predict the scalar mass spectrum:
\begin{equation}
V_{LL}=\pi^2y\Phi^4+(BL+CL^2+DL^3+EL^4)\Phi^4+ {\cdots}
\end{equation}
where $B,C,D,E$ are (dimensionless) functions of $\left(y,k,x,h\right)$, which respectively are the Higgs self-coupling $\lambda=\pi^2y$, Higgs-dark matter coupling, top quark Yukawa coupling and dark matter self-coupling; these functions have the form $\left(y^{\alpha}k^{\beta}x^{\gamma}h^{\delta}\right)L^p$ where  $p-\left(\alpha+\beta+\gamma+\delta\right)=1$ in the leading logarithm approximation.  All other Standard Model contributions such as $SU(2)$, $U(1)$ and $SU(3)$ gauge couplings are numerically sub-dominant \cite{Elias:2003xp, Chishtie:2010ni}  and have therefore been neglected. The effective potential $V_{LL}$ can be determined from the renormalization group equation  
\begin{equation}
\begin{split}
&\left(\mu\frac{\partial}{\partial\mu}+\beta^x\frac{\partial}{\partial x}+\beta^y\frac{\partial}{\partial y}+\beta^k\frac{\partial}{\partial k}+\beta^h\frac{\partial}{\partial h}\right.\\
&\left.+
\gamma_\Phi \Phi\frac{\partial}{\partial\Phi}
+\gamma_s S\frac{\partial}{\partial S}\right)V_{LL}=0\label{rg equation extension1}
\end{split}
\end{equation}
where the corresponding one loop renormalization group functions  are \cite{mura}
\begin{align}
\beta^y&=6y^2+3xy-\frac{3}{2}x^2+\frac{k^2}{128\pi^4}\label{beta1}\\
\beta^h&=3\frac{h^2}{(4\pi)^2}+12\frac{k^2}{(4\pi)^2}\label{beta2}\\
\beta^k&=4\frac{k^2}{(4\pi)^2}+3ky+\frac{kh}{(4\pi)^2}+\frac{3}{2}xk\label{beta3}\\
\beta^x&=\frac{9x^2}{4}\,,~
\gamma_\Phi=\frac{3x}{4}
\,,~\gamma_s=0
\label{gamma1}
\,.
\end{align}
Note that $\gamma_s=0$ because the $S$ field has no Yukawa coupling with Standard Model matter fields. Since we only have limited information on the renormalization group functions (to one loop order in the dark singlet model), we need to add counterterms to the effective potential to compensate for information lost due to truncation at $LL$ order
\begin{equation}
V=V_{LL}+K\left(x,y,k,h\right)\Phi^4
\end{equation}
where $K\Phi^4$ is the counterterm and $K$ is a function of the couplings. The counter terms in the full $LL$ order effective potential can be determined by the Coleman Weinberg condition \eqref{cw_condition}.

The coupling constants can be determined from the VEV conditions that spontaneous symmetry breaking will cause a nontrivial minimum in the vacuum structure:
$\frac{dV}{d\Phi}\vert_{\Phi=\mu=v}=0$.
Contact with the Standard Model is thus achieved by identifying the scale $\mu$ with the electroweak scale $\mu=v=246.2$ GeV. The mass generated for the Higgs doublet $M_H$ and dark singlet $M_S$ are only dependent on the quadratic terms in the effective potential and can be determined respectively from
\begin{equation}
M_{H}^2=\frac{dV^2}{d\Phi^2}\bigg|_{\Phi=\mu=v}\quad,\quad M_{S}^2=k\Phi^2\bigg|_{\Phi=\mu=v}\,,
\end{equation}
where we have implicitly used the result that the effective potential kinetic term renormalization constant is unity in the  Coleman-Weinberg renormalization scheme  \cite{Coleman:1973jx,Jackiw:1974cv}.

To determine the dark matter mass, Higgs mass and the corresponding Higgs coupling $y$ 
we need to input $k$ and $h$. 
However, $M_H$ shows almost no dependence on the values of these couplings in the range $0<k<1$ and $0<h<1$, with a large
suppression of $h$ contributions compared with $k$.  
At the one loop level, the predicted Higgs mass is around $216$ GeV and Higgs self-coupling is $y=0.054$ (which is $5$ times larger than the Higgs self-coupling in conventional symmetry breaking mechanism indicating the large Higgs coupling regime \cite{Wang}) for  $0<k<1$, 
in close agreement with the one loop order result given in the simplified radiative $O(4)$ model \cite{Wang,Chishtie:2010ni}.  
This implies that the singlet extension has very little effect on the Higgs mass in the considered range of $k$. This is understandable since the tree-level term $y\Phi^4$ in the Higgs mass contribution is only dependent on $y$ which makes the $k$ contribution to the Higgs mass a sub-leading loop contribution and also since $6y^2>>\frac{k^2}{128\pi^4}$ in $\beta^y$ the Higgs-dark coupling $k$ contribution is much smaller compared with the Higgs self-coupling $y$. Similarly, the dark self-coupling $h$ has an even smaller effect because it must first enter through the Higgs portal. 
Because of the small effect of the extended sector in the $O(4)$ model calculation, the radiatively-generated Higgs mass prediction can then converge to $125$ GeV when the higher loop order contributions are included \cite{Wang}. Although large values of $k$ are ruled out as dark matter solutions because of  extremely small abundances, at one loop level we reproduce the  $k\approx 6$ Higgs mass result of Ref.~\cite{Dermisek:2013pta}.

\begin{figure}[htb]
\centering
\includegraphics[width=\columnwidth]{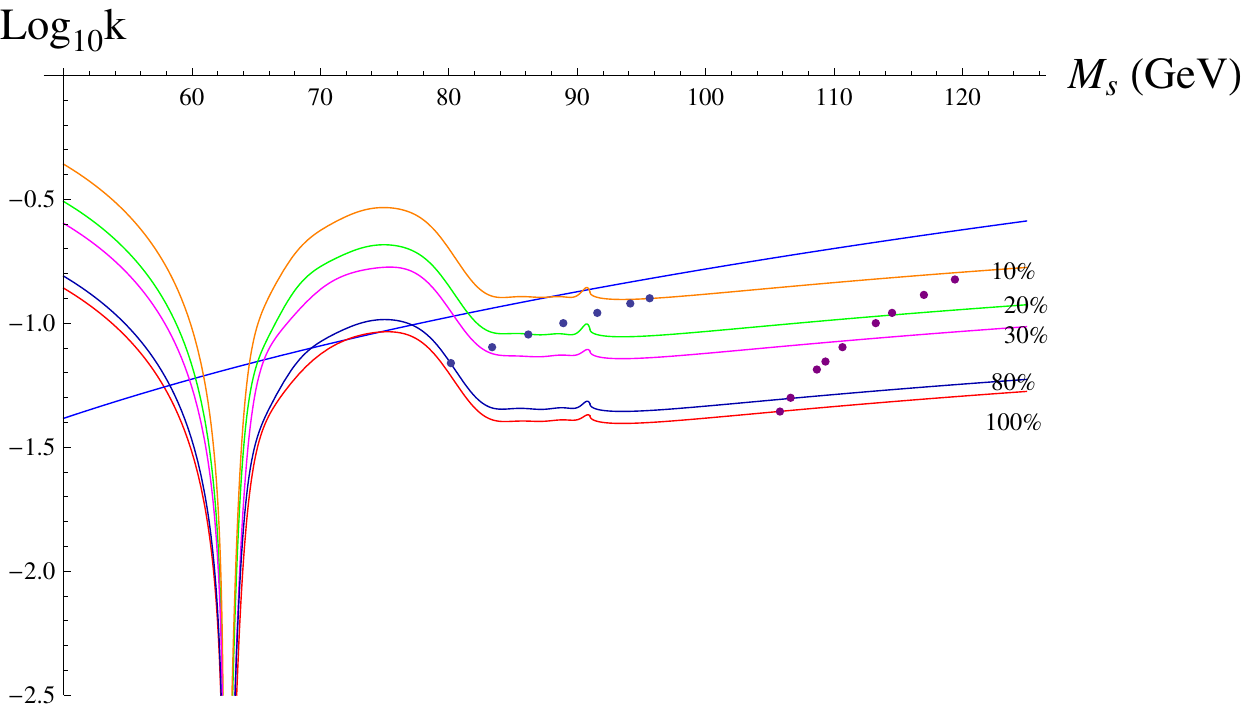}\hspace{0.05\columnwidth}
\caption{The conventional Higgs mechanism relationship between the dark matter mass and the dark-Higgs coupling (blue curve) is shown along with various dark matter abundance curves to constrain the dark singlet model.  The points correspond to the dynamical  symmetry-breaking approach for both the Higgs and dark fields at one-loop order (right set of points) and  estimated higher-loop order (left set of points).
}
\label{constraint_fig}
\end{figure}

Dark matter abundance provides a strong constraint on the dark matter mass and the corresponding dark-Higgs coupling. In Fig.~\ref{constraint_fig}, the curve of dark matter mass intersects the dark matter abundance curves corresponding to a solution for the coupling $k$ and dark matter mass $M_s$ at certain dark matter abundance. The dark matter abundance is calculated using the results of Refs.~\cite{Steigman:2012nb,Cline:ab,Dittmaier:2011ti}. However, Refs.~\cite{Cline:ab,Cline:pe} have performed a comprehensive analysis of the XENON results \cite{Aprile:2012nq} in the context of the scalar singlet model \eqref{lagrangian}, and apart from a  small region of parameter space in the $M_S\approx M_{H}/2$ resonant region, dark matter masses below $80\,{\rm GeV}$ are excluded. The resonant fine tuning region near $M_H/2$ is generally considered unnatural, and $M_S<M_H/2$ is already strongly constrained by experimental bounds on the invisible width of the Higgs \cite{Espinosa:2012im}. We thus focus on the region $M_S>80\,{\rm GeV}$ in Fig.~\ref{constraint_fig} which intersects with abundance curves below 35\%.

Thus, the sequential scenario of radiative electroweak symmetry-breaking followed by the conventional Higgs mechanism for the dark-singlet model explains less than 35\% dark matter abundance with a lower bound of $M_s>80 \,{\rm GeV}$ on the dark matter mass and $k>0.11$ on Higgs-dark matter coupling .

Consider next the alternate scenario where the  Higgs and scalar singlet masses are generated simultaneously through 
radiative electroweak symmetry breaking. Inspired  by the approach of Ref.~\cite{gildener},  in this case  the $S$ field will enter the vacuum structure along with the Higgs doublet as the fifth (gauge neutral) degree of freedom of the scalar field space. 
 However, it should be noted that though this singlet can be viewed as an extension of the scalar sector, it is different from the Higgs because it is neutral under gauge interaction and does not couple to the quarks and leptons as the Higgs doublet does. In the dynamical case, both the Higgs doublet and dark singlet enter the tree level of the effective potential and the loop corrections are of the form $L=\log\left(\frac{\Phi^2+S^2}{\mu^2}\right)$ \cite{gildener}. By contrast, in the non-dynamical case only the Higgs doublet enters the tree level effective potential, and loop corrections are of the form $L=\log\left(\frac{\Phi^2}{\mu^2}\right)$ i.e. the $S$ field does not couple directly into the logarithm. The form of the effective potential to one loop order in the dynamical case can be written as \cite{gildener}
\begin{equation}
\begin{split}
V_{LL}=&\pi^2y\Phi^4+\frac{k}{2}\Phi^2S^2+\frac{h}{24}S^2
\\&+BL+CL^2+DL^3+EL^4+\ldots
\end{split}
\end{equation}
where 
$L\equiv\log\left(\frac{\Phi^2+S^2}{\mu^2}\right)$. The quantities $B, C, D, E$ are the functions of $\left(y,k,x,h,\phi_i,S\right)$ which are dimension-4 combinations of $\Phi^2$ and $S^2$ as required by $O(4)$ and $Z_2$ symmetry and
contain leading-logarithm combinations of couplings $\left(y^{\alpha}k^{\beta}x^{\gamma}h^{\delta}\right)L^p$ where  $p-\left(\alpha+\beta+\gamma+\delta\right)=1$.
 It should be noted that $L\equiv\log\left(\frac{\Phi^2+S^2}{\mu^2}\right)$ signals that dark field $S$ affects the vacuum structure along with the Higgs field and works as a fifth scalar degree of freedom within the effective potential.
The effective potential $V_{LL}$ can be determined from the renormalization group equation which is given by Eq.~\eqref{rg equation extension1}
where the one loop renormalization group functions are the same as Eq.~\eqref{beta1}--\eqref{gamma1}. It is useful to define 
 $\rho^2=\Phi^2+S^2$ \cite{gildener}
and so truncation of the effective potential at $LL$ order
\begin{equation}
V=V_{LL}+K\left(x,y,k,h\right)\rho^4
\end{equation}
requires the $K\rho^4$ counterterm which can be determined by the Coleman Weinberg condition \cite{Coleman:1973jx,Jackiw:1974cv,Japanese}
\begin{equation}
\frac{d^4V}{d\rho^4}\bigg|_{\rho=\mu}=\frac{d^4V_{tree}}{d\rho^4}\bigg|_{\rho=\mu}
\end{equation}
where $V_{tree}$ is the tree level part of the effective potential. The vacuum structure is much more complicated in this case compared with the sequential symmetry-breaking scenario and we need two  VEV conditions consisting of one scale constraint and one nontrivial directional constraint for the minimum of the vacuum \cite{gildener}
\begin{equation}
\frac{dV}{d\rho}\bigg|_{\rho=\mu=v}=0\quad,\quad\frac{dV}{d\phi_3}\bigg|_{\stackrel{\phi_3=\mu=v}{s=0}}=0
\end{equation}
where $\phi_3$ is the component of the Higgs doublet that contains the VEV and $\mu=v=246.2$ GeV to make contact with the Standard Model. The directional constraint $\frac{dV}{dS}|_{\phi_3=\mu,s=0}=0$ is trivial since it  identically vanishes.
The dynamical mass generated for the Higgs doublet and dark singlet can be determined respectively from
\begin{equation}
M_{H}^2=\frac{dV^2}{d\phi_3^2}\bigg|_{\stackrel{\phi_3=\mu=v}{s=0}}\quad,\quad M_{S}^2=\frac{dV^2}{dS^2}\bigg|_{\stackrel{\phi_3=\mu=v}{s=0}}\,,
\end{equation}
where we have again used the result that the the effective potential kinetic term renormalization constant is unity in the  Coleman-Weinberg renormalization scheme  \cite{Coleman:1973jx,Jackiw:1974cv}.

Now we have two VEV constraints, while we have three parameters $y,k,h$ to be determined leaving one unconstrained coupling, which we choose to be $k$,  to parameterize the solutions. As discussed below, we find solutions that are perturbatively-close to $h=0$. We have chosen to parameterize our solutions through $k$ because Fig.~\ref{constraint_fig} shows that the dark matter abundance generally decreases with increasing $k$. The one-loop results are  $0.044<k<0.15$ corresponding to scalar singlet mass predictions  $106\,{\rm GeV}<M_s<120\,{\rm GeV}$  and 10\%--100\% dark matter abundance (see right-hand set of dots in Fig.~\ref{constraint_fig}). We also note that there are no one-loop leading-log solutions for $k<0.03$. The Higgs mass and Higgs self-coupling are remarkably close to the leading-log results of Ref.~\cite{Wang,Chishtie:2010ni}; hence the extended scalar sector does not destabilize radiative symmetry breaking in the Higgs sector. Comparing with the sequential symmetry-breaking scenario, the dynamical method can provide much higher levels of dark matter abundance at $LL$ order. It is interesting that the solutions lead naturally to a small scalar-singlet self interaction $h=\epsilon_1 y+\epsilon_2 k$ ($\epsilon_i\ll 1$) consistent with astrophysical evidence for weakly self-interacting dark matter \cite{bullet1,bullet2}. 
The dark matter abundance condition is surprisingly effective in constraining $k$ and the scalar singlet mass; as shown in Fig.~6 of 
Ref.~\cite{Cline:pe} the one-loop predictions are in the sensitivity region of the next generation XENON experiment.


The large Higgs self-coupling that results from the dynamical scenario  can clearly influence $M_s$ through higher loop effects.  Because the solution for $M_H$ and $y$ is very close to the radiatively-broken Standard Model result \cite{Elias:2003zm},  higher-loop effects from the Higgs portal will have a negligible effect on $M_H$ and thus the higher-loop  extrapolation to $M_H=125\,{\rm GeV}$ \cite{Wang}  will persist.  However, these higher-loop corrections from the large Higgs self-coupling could have a similar effect of decreasing the scalar singlet mass. It is important to estimate these higher-loop effects  to check if the dark matter mass  either decreases far below the $80$ GeV  lower boundary extracted from the XENON results \cite{Cline:ab,Cline:pe,Aprile:2012nq} or requires resonant fine-tuning of   $M_s$ and $k$ for acceptable dark matter abundance. 
  
The higher order estimation is based on detailed analysis of contributions to the dark matter mass from the different couplings. Because dark matter abundance constrains $k$ to be small, we assume that corrections beyond leading order to the renormalization group functions are well approximated by the $O(4)$ model. Then using the five loop results for the $O(4)$ renormalization group functions \cite{Kleinert:1991rg} combined with the extrapolation methods of \cite{Wang}, we obtain the following result for the dark matter mass:
\begin{equation}
\begin{split}
M_s^2&=-1461.56\left(\frac{k}{0.05}\right)\left(\frac{y}{0.0534}\right)
\\&+3025.8\left(\frac{k}{0.05}\right)
+100^2\left(\frac{y}{0.0534}\right)^{1.4}\,,\label{empirical mass formula}
\end{split}
\end{equation}
where we are working in GeV units and the top quark Yukawa coupling contributions are embedded in the numerical coefficients. Only the dominant leading order contributions in $k$ have been retained in \eqref{empirical mass formula} (i.e. higher order terms in $k$ are numerically suppressed because $k\sim0.1$). By using this formula, we can estimate the dark matter mass at the convergence value $y=0.0233, M_{Higgs}=125$ GeV of Ref.~\cite{Wang}. As discussed earlier, we use the lower-bound $M_s>80\,{\rm GeV}$ which is the lowest mass consistent with analysis of the  XENON results \cite{Cline:ab,Cline:pe,Aprile:2012nq} without $M_S\approx M_H/2$ resonance fine-tuning. The higher order estimation gives the dark-Higgs coupling $0.07<k<0.13$ for dark matter mass $80\,{\rm GeV}<M_s<96\,{\rm GeV}$, and dark matter abundance 10\%--80\% as shown by the left-hand set of dots in Fig.~\ref{constraint_fig}. In general,  the estimated higher-loop effects result in a significant reduction of the dark matter mass compared to the one-loop predictions for comparable levels of abundance.

We have also studied the possibility of spontaneous breaking of the $Z_2$ symmetry by allowing a non-zero rotation angle in the VEV 
(i.e.,  $\langle S\rangle=v\sin\theta$,  $\langle \Phi\rangle=v\cos\theta$) and self-consistently determining the couplings in each case using the procedure outlined above.  For physical solutions of the couplings, the  vacuum energy of the $Z_2$-symmetric case is  always found to be smaller, providing evidence that $Z_2$ symmetry remains unbroken. 

We have studied radiative symmetry breaking in an extension of the conformally invariant Standard Model containing a scalar singlet field with a Higgs portal interaction.  The sequential symmetry-breaking scenario, where electroweak symmetry-breaking occurs via a large Higgs self coupling and the scalar singlet mass is then generated by the conventional Higgs mechanism, can explain at most 35\% of the dark matter abundance without resonant fine-tuning.  By contrast, the dynamical approach inspired by Ref.~\cite{gildener}, where the electroweak and the $Z_2$-symmetric scalar-singlet vacuum simultaneously result from radiative symmetry-breaking in the large Higgs-coupling regime can accommodate larger dark matter abundances and results in a weakly self-interacting scalar singlet.  Estimating the higher-loop effects needed to  maintain  consistency with a radiatively-generated 125 GeV Higgs mass dominated by the large Higgs self-coupling leads to the bounds $80\,{\rm GeV}<M_s<96\,{\rm GeV}$ and a corresponding dark matter abundance in the range 10\%--80\%.  The dark matter mass and Higgs-portal coupling predictions of the dynamical scenario, both at one-loop and estimated higher-loop levels, are within the range of sensitivity of the next generation of the XENON experiment \cite{Cline:ab,Cline:pe}.

T.G.S  and R.B.M are grateful for financial support from the Natural Sciences and Engineering Research Council of Canada (NSERC).  We thank Cliff Burgess, Maxim Pospelov and Rainer Dick for helpful suggestions.

\end{document}